\documentclass[reprint, aps, prd, nofootinbib, table]{revtex4-1}

\usepackage{subcaption, ragged2e}
\DeclareCaptionJustification{justified}{\justifying}
\usepackage{graphicx}
\usepackage{amsmath, amssymb, amsbsy, amstext, amsthm, simplewick, amsfonts}
\usepackage{dcolumn}
\usepackage{mathtools}
\usepackage{enumitem}
\usepackage{bm}
\usepackage{float}
\usepackage[many]{tcolorbox}
\usepackage{shuffle}
\usepackage{environ}
\usepackage{physics}
\usepackage{tabularx}
\usepackage{diagbox}
\usepackage{multirow}
\usepackage{braket}

\usepackage{hyperref}
\hypersetup{
    colorlinks=true,
    linkcolor={rosewood},
    citecolor={rosewood},
    urlcolor={rosewood}
}

\definecolor{pyblue}{RGB}{31, 119, 180}
\definecolor{pyorange}{RGB}{255, 127, 14}
\definecolor{pygreen}{RGB}{44, 160, 44}
\definecolor{pyred}{RGB}{214, 39, 40}
\definecolor{lightgray}{gray}{0.9}
\definecolor{rosewood}{rgb}{0.4, 0.0, 0.04}

\usepackage{tikz}
\usetikzlibrary{calc}
\usetikzlibrary{fadings}
\usetikzlibrary{arrows,calc}
\usetikzlibrary{decorations.pathmorphing}
\tikzstyle{BD}=[thick,line cap=round,black,decorate,decoration={
	snake,amplitude=.4mm,segment length=2.5mm,post length=1mm}]

\def \d {\mathrm{d}}

\def \C {\mathcal{C}}
\def \D {\mathcal{D}}

\def \L {\mathcal{L}}
\def \M {\mathcal{M}}

\def \O {\mathcal{O}}
\def \P {\mathcal{P}}

\def \fnl {f_{\text{NL}}}

\def \Mpl {M_{\text{pl}}}

\begin{document}

\title{Universal Non-Gaussian Signatures from Transient Instabilities}

\author{Shuntaro Aoki$^{1}$}
\email{shuntaro.aoki@riken.jp}

\author{Diederik Roest$^{2}$}
\email{d.roest@rug.nl}

\author{Denis Werth$^{3, 4, 5}$}
\email{denis.werth@mpp.mpg.de}

\affiliation{\vskip 4pt $^{1}$RIKEN Center for Interdisciplinary Theoretical and Mathematical Sciences, Wako, J-351-0198, Japan \\
$^{2}$Van Swinderen Institute for Particle Physics and Gravity, University of Groningen, Groningen, NL-9747, The Netherlands \\
$^{3}$Institut d’Astrophysique de Paris, Sorbonne Université, CNRS, Paris, F-75014, France \\
$^{4}$Max Planck Institute for Physics, Werner-Heisenberg-Institut, Munich, D-85748, Germany \\
$^{5}$Max Planck-IAS-NTU Center for Particle Physics, Cosmology and Geometry}

\begin{abstract}
We identify universal signatures in the bispectrum arising from a transient tachyonic instability of entropic fluctuations during inflation, a phenomenon that naturally arises in hyperbolic field-space geometries. We perform exact numerical calculations directly at the level of fluctuations, without relying on a specific background model, and distinguish two cases. In the light case, with masses around the Hubble scale, our results provide the first-ever computation of the bispectrum due to such tachyonic instabilities. We find a universal magnification of the folded configuration, together with the known non-analytic scaling in the squeezed limit.  As an illustrative example, we compute and analyze the bispectrum in angular inflation, demonstrating compatibility with current limits. In the heavy case, with masses well above the Hubble scale, the bispectrum exhibits  a distinctive correlation between magnified folded configurations and a `tachyonic resonance' in mildly squeezed limits, with the resonance scale set by the strength of the instability. While the main qualitative features are reproduced, we show that there exists no UV matching for which a single-field effective description, obtained by integrating out the entropic modes, accurately captures the bispectrum for all kinematic configurations. To facilitate observational applications, we introduce simple bispectrum shape templates suitable for current and forthcoming cosmological surveys. Our model-independent results allow for constraining non-standard inflationary attractors characterized by strongly non-geodesic motion.
\end{abstract}

\maketitle


\section{Introduction}

\noindent
Initially proposed to explain the background evolution of the Universe, cosmic inflation now draws most of its observational punch from the study of fluctuations. Measurements of the cosmic microwave background have placed stringent bounds on the nearly scale-invariant primordial power spectrum at large scales, constraining the scalar spectral tilt, $n_s=0.974\pm 0.003$, through the latest joined ACT/DESI/Planck data~\cite{Planck:2018jri, ACT:2025fju, DESI:2024mwx}. Moreover, the absence of a detected primordial gravitational wave signal translates into an upper bound on the tensor-to-scalar ratio $r$. In the simplest single-field slow-roll scenario, this restricts the inflaton potential and requires a small inflaton velocity (in Planck units). This provides a clear example of how fluctuation observables can be used to constrain background dynamics.

\vskip 4pt
However, embedding inflation within a more fundamental high-energy framework complicates this simple picture: additional fields are generically expected as, for instance, moduli arising in string compactifications~\cite{Linde:1990flp, Baumann:2014nda}. These extra degrees of freedom can substantially alter observable predictions as they naturally lead to richer background dynamics, with e.g.~curved field spaces and strongly turning trajectories. These can be realized in steep regions of the string landscape potential, a scenario that has gained much attention in light of the recently discussed `swampland criteria', which bound field excursions in field space and constrain the slope of the potential~\cite{Arkani-Hamed:2006emk, Obied:2018sgi, Garg:2018reu, Ooguri:2018wrx, Agrawal:2018own, Danielsson:2018ztv, Achucarro:2018vey, Bravo:2020wdr}. In this context, the mapping between a handful of well-constrained observables and the many parameters characterizing a multi-field potential and curved field space becomes both delicate and highly degenerate. Before tackling these ambitious questions, however, a key step is to first identify the very nature of the inflationary background itself, in particular, whether it is fundamentally multi-field. 

\vskip 4pt
Recently, considerable attention has been devoted to negatively curved field spaces, which naturally arise in the low-energy limits of supergravity models. These can give rise to $\alpha$-attractors, whose predictions are insensitive to the precise form of the inflationary potential~\cite{Kallosh:2013yoa, Carrasco:2015uma, Achucarro:2017ing, Linde:2018hmx}. Alternatively, slow-roll trajectories in such geometries can be prone to geometrical destabilization~\cite{Gong:2011uw, Renaux-Petel:2015mga, Renaux-Petel:2017dia, Cicoli:2018ccr, Grocholski:2019mot}, after which they are attracted towards alternative solutions characterized by strongly non-geodesic motion~\cite{Cremonini:2010ua, Brown:2017osf, Mizuno:2017idt, Garcia-Saenz:2018ifx, Bjorkmo:2019aev, Christodoulidis:2019mkj, Christodoulidis:2019jsx, Aragam:2019omo}. This generic and broad class of `slow-roll fast-turn' attractors is defined by a large (compared to slow-roll parameters) slowly varying multi-field dimensionless turn rate $\eta_\perp$ that allows for a stable and prolonged phase of inflation. These models share the following criteria~\cite{Bjorkmo:2019fls}
\begin{equation}
    \epsilon, |\eta| \ll 1\,, \quad \eta_\perp^2 \gg \O(\epsilon)\,, \quad \frac{\dot{\eta}_\perp}{H\eta_\perp} \ll 1\,, \label{slow-turn}
\end{equation}
where $\epsilon$ and $\eta$ are the first and second slow-roll parameters, respectively, and a dot denotes derivative with respect to cosmic time, $\dot{f} \equiv \tfrac{\d f}{\d t}$. 

\vskip 4pt
In this paper we will focus on this regime, as a strong non-geodesic motion (i.e.~the large turn rate  $\eta_\perp$) naturally gives rise to a transient tachyonic instability of entropic fluctuations. As we will discuss, these give rise to distinctive and universal signatures in the bispectrum shape - applicable to all slowly-varying backgrounds of the form \eqref{slow-turn}.\footnote{See e.g.~\cite{Achucarro:2022qrl} for a general review on primordial non-Gaussianities and the references therein.} Our analysis is performed directly at the level of fluctuations, where for concreteness we focus on two fields, and remains completely agnostic about any specific background, thereby ensuring full generality. Employing the {\sf CosmoFlow} code~\cite{Werth:2023pfl, Pinol:2023oux, Werth:2024aui}, we perform an exact numerical computation of the bispectrum shapes generated from intrinsically multi-field cubic interactions, and identify the following features shared by all cubic operators that are not slow-roll suppressed. Importantly, as we assume the time-dependence of background quantities to be slow-roll suppressed, the bispectrum will be scale-invariant (allowing for a more efficient comparison to data, see e.g.~\cite{Philcox:2026njr}).

\vskip 4pt
Let us first outline the results for heavy isocurvature fluctuations, where the entropic mass is (much) larger than the Hubble parameter. In addition to an oscillatory pattern in the soft limit, the bispectrum shapes display a correlation between a magnified signal in the folded kinematic configuration compared to the equilateral limit, and a characteristic `tachyonic resonance' in mildly-soft configurations. The position of this resonance is directly related to the strength of the instability. 

\vskip 4pt
It has previously been shown that these distinctive signatures of a transient tachyonic instability with a large mass can be effectively captured within a single-field description, in which the curvature perturbation modes propagate with an imaginary speed of sound~\cite{Garcia-Saenz:2018ifx, Garcia-Saenz:2018vqf, Fumagalli:2019noh, Bjorkmo:2019qno, Ferreira:2020qkf, Garcia-Saenz:2025jis, Iarygina:2023msy}. A detailed comparison shows that, although this effective theory describes the main qualitative behavior well, it does not reproduce the bispectrum shape fully accurately in kinematic configurations that feature a (modest) hierarchy of scales. Indeed, we find that the precise UV matching of the effective parameters depends on the kinematics, highlighting the limitations of such a description in the presence of different scales. This implies that the single-field description does not capture the full bispectrum shape across all physical kinematic configurations, and that a genuine multi-field calculation is required.

\vskip 4pt
In addition to the heavy case, our exact multi-field approach allows one to consider inflationary scenarios where both the entropic mass and the mild trajectory’s bending rate are of order unity or smaller in Hubble units. These intermediate regimes are not captured by the simplified single-field effective description, and again we find a stronger signal in the folded limit compared to the equilateral configuration.  As an example of this case, we focus on a specific and concrete background, provided by the model of angular inflation~\cite{Christodoulidis:2018qdw}, which exhibits a moderate degree of turning. We compute its exact bispectrum,  and show that it displays all the identified signatures of transient instability of the entropic mode.

\vskip 4pt
To facilitate phenomenological applications, we propose ready-to-use bispectrum shape templates that capture all these signatures. Observing these signatures would provide compelling evidence for, and strongly point towards, a non-geodesic motion in the underlying field-space geometry.

\vskip 4pt
The outline of the paper is as follows: In Sec.~\ref{sec: multi-field fluctuations with transient instabilities}, we introduce the model directly at the level of fluctuations and discuss destabilization of entropic modes, leading to a transient tachyonic instability. In Sec.~\ref{sec: Hyperbolic shapes} we perform an exact numerical calculation of the resulting non-Gaussianities and show that they exhibit distinctive and unique features. We then compare these properties with a previously-studied single-field effective field theory (EFT) and discuss its limitations. Motivated by our findings, we introduce a new family of non-geodesic shape templates that accurately captures the genuine multi-field signatures identified in our analysis. In Sec.~\ref{sec: Concrete realization}, we derive a universal bound on background quantities that generically leads to a transient tachyonic instability, and compute the bispectrum in angular inflation. We show that it exhibits all the predicted features and is well described by the proposed template. Finally, our conclusions are stated in Sec.~\ref{sec: conclusions}. App.~\ref{app} offers a comparison of the super-horizon and single-field limits.

\section{multi-field fluctuations with transient instabilities}
\label{sec: multi-field fluctuations with transient instabilities}

\noindent
We first introduce the model and discuss the generic feature of a transient instability of entropic modes.

\subsection{Background dynamics}

\noindent
Our focus will be on inflationary models that comprise the general class of nonlinear sigma models---a set of scalar fields with non-canonical kinetic term---minimally coupled to gravity. The action is given by
\begin{equation}
    S = \int \d^4x\sqrt{-g}\left[  \tfrac{1}{2} {\Mpl^2} R -  \tfrac{1}{2} G_{IJ}\nabla_\mu \phi^I\nabla^\mu \phi^J - V\right]\,,
\end{equation}
where $\phi^I$ ($I=1, \ldots, N$) are $N$ scalar fields, $V$ is the multi-field potential and $G_{IJ}$ is the internal field-space metric that will play a crucial role in what follows. Moreover, $R$ denotes the Ricci scalar constructed out of the spacetime metric with determinant $g$. The inflationary background is characterized by the evolution of the homogeneous fields $\bar{\phi}^I(t)$ and a quasi-de Sitter background with scale factor $a(t)$.

\vskip 4pt
For concreteness we will specialize to two-field models. In that case, the background evolution defines two orthogonal directions, tangent and normal to the background trajectory~\cite{Gordon:2000hv, GrootNibbelink:2001qt}:
\begin{align}
     t^I = \dot{\bar{\phi}}^I/\dot{\bar{\phi}} \,, \quad
     n_I = |G|^{1/2}\epsilon_{IJ}t^J \,,
\end{align}
where $\dot{\bar{\phi}}^2 \equiv G_{IJ}\dot{\bar{\phi}}^I\dot{\bar{\phi}}^J$ denotes the total background velocity and $\epsilon_{IJ}$ is the Levi-Civita symbol. An important parameter will be the turn rate $\eta_\perp$, defined by
\begin{equation}
\label{eq: turn rate def}
    \D_t t^I = -H \eta_\perp n^I \,,
\end{equation}
where $\D_t$ is the covariant (cosmic) time derivative, whose action on a given vector $A^I$ is such that $\D_tA^I\equiv \dot{A}^I + \Gamma^I_{JK}\dot{\bar{\phi}}^JA^K$, where $\Gamma^I_{JK}$ are the Christoffel symbols derived from $G_{IJ}$. This dimensionless parameter measures the deviation of the background trajectory from a field-space geodesic.

\subsection{Linear perturbations}

\noindent
Fluctuations are naturally defined along the tangent and normal directions. In particular, curvature and isocurvature perturbations are defined by 
\begin{equation}
    \zeta = -t_I\delta\phi^I\, H/\dot{\bar{\phi}} \,, \quad \sigma = n_I \delta \phi^I \,,
\end{equation}
where $\delta\phi^I \equiv \phi^I - \bar{\phi}^I$. At the linear level, their dynamics is governed by the quadratic action
\begin{equation}
    \L^{(2)}/a^3 = -\Mpl^2\epsilon\,(\partial_\mu \zeta)^2  - \tfrac{1}{2}(\partial_\mu \sigma)^2 - \tfrac{1}{2} m_\sigma^2 \sigma^2 + 2\dot{\bar{\phi}}\,\eta_\perp \dot{\zeta}\sigma \,. \label{quadratic}
\end{equation}
The free theory is therefore fully specified by two parameters. The first is the (bare) mass of the isocurvature (or entropic) fluctuation $\sigma$, given by
\begin{equation}
    m_\sigma^2 = V_{NN} - \eta_\perp^2H^2 +\epsilon H^2\Mpl^2 R_{\rm fs}\,,
\end{equation}
where $V_{NN} \equiv n^I n^J \D_I\D_J V$ and $R_{\rm fs}$ is the field-space scalar curvature. The second is the turn rate $\eta_\perp$, defined in~\eqref{eq: turn rate def}: non-geodesic motion therefore couples curvature and isocurvature modes. 

\vskip 4pt
Note that a negative bare entropic mass can arise naturally e.g.~due to a large turn rate, given its negative definite contribution. Moreover, it follows from the last term that this effect is naturally enhanced by a negative field-space curvature. Indeed, we will argue in Sec.~\ref{sec: bound on background dynamics} that such geometry-induced instability is a common feature of hyperbolic field-space geometries by deriving a universal bound on background parameters leading to a transient tachyonic instability.


\begin{figure*}[t]
    \centering
    \begin{subfigure}{0.32\textwidth}
        \centering
        \includegraphics[width=\textwidth]{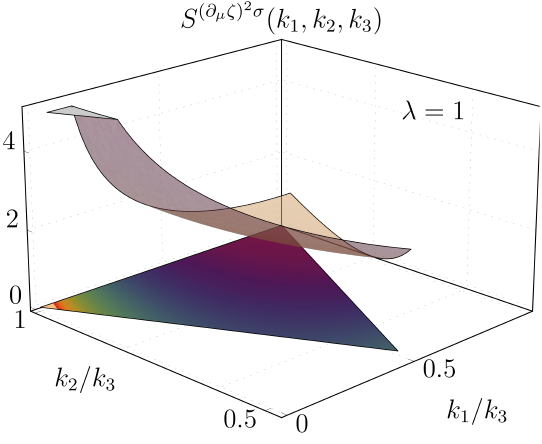}
    \end{subfigure}\hfill
    \begin{subfigure}{0.32\textwidth}
        \centering
        \includegraphics[width=\textwidth]{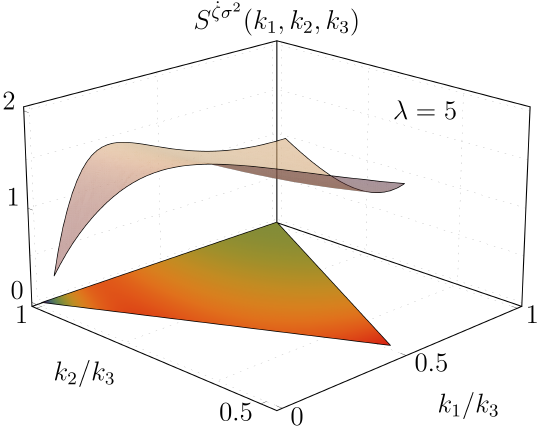}
    \end{subfigure}\hfill
    \begin{subfigure}{0.32\textwidth}
        \centering
        \includegraphics[width=\textwidth]{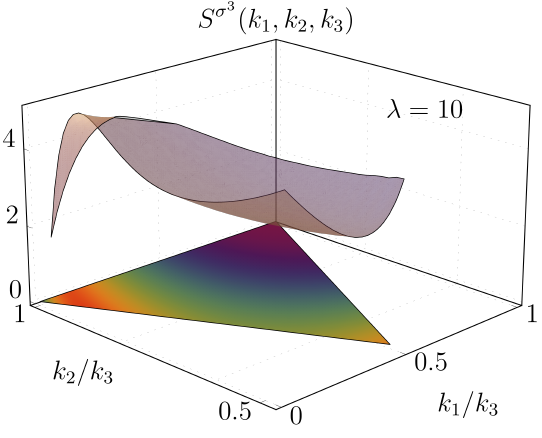}
    \end{subfigure}
    \caption{\it Dimensionless bispectrum shapes $S(k_1, k_2, k_3)$, normalized to unity in the equilateral configuration $k_1=k_2=k_3$, in the entire kinematic configurations, induced by the interactions $(\partial_\mu \zeta)^2\sigma$ for $\lambda=1$ ({\rm left panel}), $\dot{\zeta}\sigma^2$ for $\lambda=5$ ({\rm middle panel}), and $\sigma^3$ for $\lambda=10$ ({\rm right panel}).
    }
    \label{fig: hyperbolic shape examples}
\end{figure*}

\vskip 4pt
Interestingly, even in the presence of such an instability, it can happen that the background remains stable---the saving grace in such cases is the mixing with the curvature mode, ensuring that the instability is localized in time. In particular, the primary effect of non-geodesic motion is to generate an effective entropic mass (see e.g.~\cite{Achucarro:2010da})
\begin{equation}
    m_{\sigma, \text{eff}}^2 = m_\sigma^2 + 4\eta_\perp^2H^2\,,
\end{equation}
that governs superhorizon evolution. Note that the turn rate now contributes positively, thus allowing for a positive effective mass:\footnote{Backgrounds with a negative effective mass can still be stable in the presence of a rapidly decreasing  turn rate, see e.g.~\cite{Achucarro:2026qyx} - instead, we assume the turn rate to be constant.}
\begin{equation}
    m_\sigma^2 < 0 < m_{\sigma, \text{eff}}^2 \,.
\end{equation}
For simplicity, we choose the following relation between the bare and effective entropic masses:\footnote{Notice that this parametrization is precisely realized in hyper-inflation~\cite{Brown:2017osf}.}
\begin{equation}
\label{eq: masses parametrization}
    m_\sigma^2 = -\lambda^2 H^2\,, \quad m_{\sigma, \text{eff}}^2 = +\lambda^2 H^2\,,
\end{equation}
where $\lambda>0$ is positive real dimensionless (constant) number. This choice collapses the two free parameters of the linear theory to a single one, related to the turn rate via $\eta_\perp = \lambda/\sqrt{2}$. In App.~\ref{app2}, we discuss our results for the case in which $m_\sigma^2$ and $m_{\sigma,\rm{eff}}^2$ are parameterized by two independent parameters.

\vskip 4pt
The above mass assignment defines the transient instability that we focus on in this paper. As a consequence, isocurvature modes undergo an instability just after mass-shell crossing of the $\sigma_k$ mode, $k/a \sim |m_\sigma|$ where $k$ is the Fourier wavenumber, while on super-horizon scales, $k/a \ll H$, these modes are stable and decay. The resulting growth of entropic fluctuations feeding the curvature perturbation sector through the linear mixing results in an enhancement of the primordial power spectrum and an exponentially small tensor-to-scalar ratio.\footnote{It has been shown in~\cite{Garcia-Saenz:2025jis} that the exponential growth of the scalar fluctuations is transferred to the tensorial sector through loop corrections.} When the entropic field decays, the curvature perturbation eventually freezes out.

\subsection{Non-linear interactions}

\noindent
The transient growth of isocurvature fluctuations allows for the generation of sizeable non-Gaussianities in the presence of non-linear interactions. 
Neglecting slow-roll suppressed terms,\footnote{Non-shift symmetric slow-roll suppressed interactions are parametrically smaller than interactions $\propto \eta_\perp^n$ with $n \geq1$.} the leading (hence shift-symmetric in $\zeta$) cubic interactions in non-linear sigma models are~\cite{Garcia-Saenz:2019njm}
\begin{equation}
\label{eq: cubic interactions}
    \begin{aligned}
        \L^{(3)}_{\zeta^2\sigma}/a^3 &= \tfrac{\dot{\bar{\phi}}}{H}\, \eta_\perp(\partial_\mu \zeta)^2\sigma\,, \\
        \L^{(3)}_{\zeta\sigma^2}/a^3 &= \tfrac{1}{H}(\epsilon \Mpl^2H^2 R_{\rm fs}-\eta_\perp^2H^2)\,\dot{\zeta}\sigma^2\,, \\
        \L^{(3)}_{\sigma^3}/a^3 &= -\tfrac{1}{6}(V_{NNN} - 2\dot{\bar{\phi}}H \eta_\perp R_{\rm fs})\,\sigma^3\,,
    \end{aligned}    
\end{equation}
which, at weak mixing $\eta_\perp\ll 1$, lead to single-, double-, and triple-exchange diagrams for the bispectrum, respectively. 

\vskip 4pt
Being not slow-roll suppressed, these interactions can generically lead to parametrically large non-Gaussian signals, of order $\fnl=\O(1)$ in equilateral configurations; see~\cite{Pinol:2023oux} for a detailed study of perturbativity bounds on the corresponding Wilson coefficients.

\section{Bispectrum shapes}
\label{sec: Hyperbolic shapes}

\noindent
We now turn to the study of new signatures on the non-Gaussianity phenomenology due to the transient tachyonic instability. To this end, we compute the resulting dimensionless bispectrum shape, defined as 
\begin{equation}
    S(k_1, k_2, k_3) \equiv \frac{(k_1 k_2 k_3)^2}{(2\pi)^4\P_\zeta^2} \braket{\zeta_{\bm{k}_1} \zeta_{\bm{k}_2} \zeta_{\bm{k}_3}}' \,,
\end{equation}
where a prime denotes that we have stripped off an overall momentum-conserving delta function, and $\P_\zeta = 2.2\times10^{-9}$ is the dimensionless primordial power spectrum. Moreover, we will normalize these shapes to unity in the equilateral limit, i.e.~$S(k, k, k)=1$, thus effectively absorbing the Wilson coefficients of the cubic interactions~\eqref{eq: cubic interactions} into this definition. The resulting shapes therefore only depend on the choice of cubic operator and the mass scale $\lambda$. These shapes have been computed numerically using {\sf CosmoFlow}~\cite{Werth:2023pfl, Pinol:2023oux, Werth:2024aui}, which accounts for linear mixings in a fully non-perturbative manner, as required for $\lambda \geq 1$. We neglect any variation of the Hubble parameter, therefore assuming perfect scale-invariance, in which case slow-roll suppressed cubic interactions (including self-interactions, e.g.~$\dot{\zeta}^2\zeta$) vanish.

\vskip 4pt
We include a number of examples of the resulting shapes in Fig.~\ref{fig: hyperbolic shape examples}, where we show the dimensionless bispectrum shapes in all kinematic configurations, for a selected cubic interaction and for three representative values of the strength of the instability: $\lambda=1$ illustrates the light case, and $\lambda=5$ and $\lambda=10$ are two examples of the heavy case. Moreover, in Fig.~\ref{fig: Shape slices}, we show the bispectrum shapes for isosceles kinematic configurations (where $k_2 = k_3$) for all cubic interactions, again for the same three representative mass values. As can be seen from these figures, the main features are largely independent of the choice of cubic interaction, with all three leading to similar results. These shapes exhibit several distinctive features:
\begin{itemize}
    \item Firstly, they display the standard {\it cosmological collider signal} in the squeezed limit $k_1 \ll k_2 \sim k_3$, determined by the effective entropic mass $m_{\sigma,\text{eff}}$;
    \item Second, they show a {\it magnified folded limit} $k_1 \sim k_2 \sim k_3/2$, which is particularly pronounced for shapes generated by the interaction $\L^{(3)}_{\zeta^2\sigma}$;
    \item Finally, in the case of a heavy entropic field, we observe a {\it characteristic resonance} in mildly soft kinematic configurations with $k_1 \lesssim k_2 \sim k_3$, most clearly visible in the right panel of Fig.~\ref{fig: hyperbolic shape examples}.
\end{itemize}
We now discuss each of these features in turn.

\subsection{Soft behavior in squeezed limit}

\noindent
Soft limits of cosmological correlators are well known to encode information about additional field content. If the curvature perturbation couples to a heavy field with mass much larger than the Hubble scale ($m \gg H$), the latter decays rapidly on super-horizon scales, producing an almost equilateral shape that vanishes in the squeezed limit, $S \sim k_1/k_3$~\cite{Maldacena:2002vr}. In the opposite regime of a very light field ($m \ll H$), the additional mode is long-lived and continuously sources the observable sector, leading to an enhanced squeezed limit. In the limiting case $m/H \to 0$, this reproduces the standard local shape $S \sim k_3/k_1$. For intermediate masses $m \sim H$, the shapes display the characteristic non-analytic scaling~\cite{Noumi:2012vr, Chen:2009we, Chen:2009zp, Arkani-Hamed:2015bza}
 \begin{align}
  S \sim (k_3/k_1)(k_1/k_3)^{3/2-\nu_\lambda} \,, \quad \nu_\lambda \equiv \sqrt{9/4-\lambda^2} \,.
 \end{align}
This scaling reflects the super-horizon decay governed by the effective entropic mass. Depending on the value of $\lambda$, the scaling can be of power-law type ($\lambda\leq 3/2$) or oscillatory ($\lambda\geq 3/2$); these correspond to the complementary and principal series of de Sitter unitary irreducible representations, respectively. Importantly, this signal is an intrinsic multi-field effect. 

\begin{figure}[t!]
    \centering
    \includegraphics[width=7.5cm]{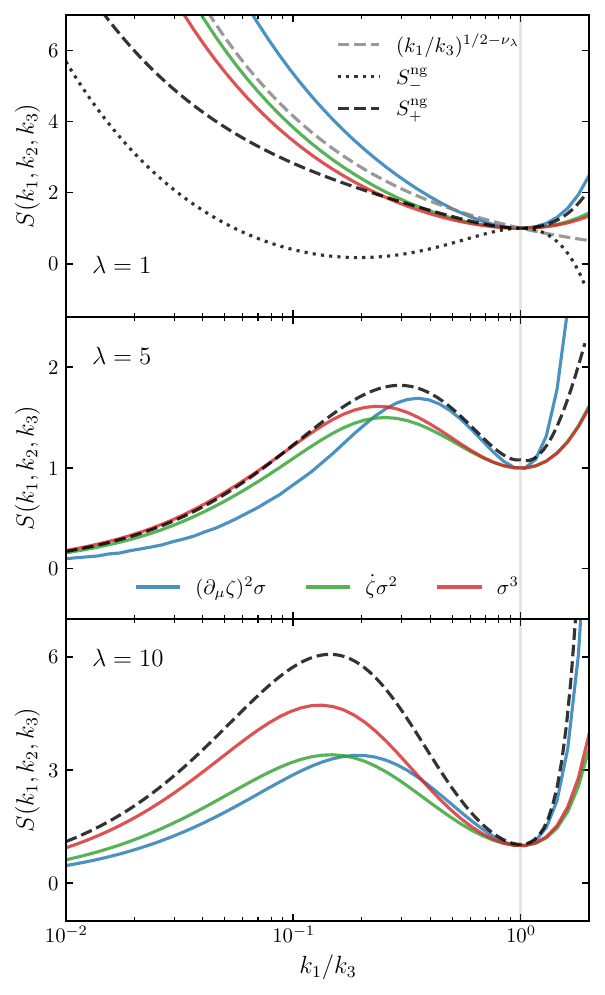}
    \caption{\it Slices of the dimensionless bispectrum shapes $S(k_1, k_2, k_3)$ in the isosceles configuration $k_2=k_3$ from the folded limit $k_1/k_3=2$ to the squeezed limit $k_1/k_3\to0$, for each interactions given in~Eq.~\eqref{eq: cubic interactions}, for $\lambda=1$ ({\it upper panel}), $\lambda=5$ ({\rm middle panel}) and $\lambda=10$ ({\rm lower panel}). We have normalized the shapes to unity in the equilateral configuration, represented by the vertical gray line. The typical power-law scaling $S\sim (k_1/k_3)^{1/2-\nu_\lambda}$ for the light entropic field is represented by the gray dashed line. We show the non-geodesic shape templates $S^{\rm ng}_\pm$, defined in~\eqref{eq: light hyperbolic template} and~\eqref{eq: heavy hyperbolic template}, in black dashed and dotted lines. The dotted $S_-^{\rm ng}$ curve reproduces the shape predicted by angular inflation, see Sec.~\ref{subsec: bispectrum shape}.}
    \label{fig: Shape slices}
    \vspace{-1cm}
\end{figure}

\vskip 4pt
In Fig.~\ref{fig: Shape slices}, the power-law scaling is clearly visible for $\lambda=1$. However, for a heavy entropic field ($\lambda=5$ and $\lambda=10$) the oscillations only show up at much smaller scales, namely for $k_1/k_3 \ll 10^{-2}$, not visible in Fig.~\ref{fig: hyperbolic shape examples}. This kinematic region only corresponds to a small portion of the full kinematically allowed space.

\subsection{Magnification in the folded limit}

\noindent
Moving away from the equilateral configuration into the folded limit, with $k_1 \sim k_2 \sim k_3/2$, the signal in the bispectrum shape increases; this applies to all three cubic interactions and for all mass values, and is best seen in Fig.~\ref{fig: Shape slices}. This strong signal is a direct consequence of the transient tachyonic instability in the entropic $\sigma$-sector being transferred to the observable $\zeta$-sector through the linear mixing and cubic interactions.
This is analogous to models of inflation that already start in an excited quantum state where a folded-limit magnification arises from the mixing of positive- and negative-frequency modes~\cite{Chen:2006nt, Holman:2007na, Meerburg:2009ys}.

\vskip 4pt
In the large entropic mass limit ($\lambda\gg1$), this feature can be understood within an effective single-field description~\cite{Garcia-Saenz:2018vqf, Fumagalli:2019noh, Bjorkmo:2019qno}. Integrating out the heavy field yields an effective single-field EFT with an imaginary speed of sound. For later convenience, it is useful to define the canonically normalized curvature perturbation field $\zeta_c$ such that $\zeta_c^2 \equiv 2 \Mpl^2 \epsilon \zeta^2 = \dot{\bar{\phi}}^2/H^2 \zeta^2$. The multi-field quadratic theory reads
\begin{equation}
    \L^{(2)}/a^3 = \tfrac{1}{2}\zeta_c\Box \zeta_c + \tfrac{1}{2}\sigma(\Box-m_\sigma^2)\sigma + 2 \eta_\perp H \dot{\zeta}_c\sigma \,,
\end{equation}
where we have omitted the more slow-roll–suppressed terms. In the heavy entropic field limit, the equation of motion for $\sigma$ becomes $m_\sigma^2 \sigma = 2\eta_\perp H \dot{\zeta}_c$ (see Appendix~\ref{app} for a discussion of the implications of the heavy vs.~the superhorizon limit). Substituting this into the quadratic action yields the effective single-field EFT with speed of sound~\cite{Garcia-Saenz:2018vqf}
\begin{equation}
    c_s^{-2}\equiv m_{\sigma, \text{eff}}^2/m_\sigma^2<0 \,.
\end{equation}
In the regime with a transient instability, $c_s$---which retains the remaining physical information regarding the entropic mode---is therefore imaginary, $c_s = i|c_s|$. For our choice~\eqref{eq: masses parametrization}, we simply have $c_s = i$. 

\vskip 4pt
This effective single-field description leads to the following mode function for the curvature perturbation
\begin{equation}
\label{eq: EFT mode function}
    \begin{aligned}
        \zeta_{c, k}(\tau) = \frac{\alpha_k}{k^{3/2}}&\left[(1+|c_s|k\tau)e^{-|c_s|k\tau} \right.\\
        &\left.+ \rho_ke^{i\theta_k}(1-|c_s|k\tau)e^{+|c_s|k\tau}\right]\,,
    \end{aligned}
\end{equation}
where $\tau$ is conformal time, defined as $\d\tau=\d t/a$, and $\alpha_k, \rho_k$ and $\theta_k$ are positive real coefficients whose mild $k$-dependence will be neglected for simplicity. The non-standard quantization condition $\alpha_k^2 \rho_k\sin(\theta_k) = H^2/8\epsilon|c_s|\Mpl^2$ fixes the power spectrum amplitude, and enforces a non-vanishing and out of phase growing mode, i.e.~$\rho_k\neq0$ and $\theta_k\neq0$ (mod $2\pi$). 

\vskip 4pt
Computing the bispectrum would mix the exponentially growing and decaying modes in~Eq.~\eqref{eq: EFT mode function}. The momentum dependence of the bispectrum can then be easily understood. When all three growing (internal) modes interact constructively, the time integral, which schematically reads $\int^0\d\tau \, \tau^n e^{(k_1+k_2+k_3)|c_s|\tau}$, generates equilateral-like shapes, where $n$ is interaction dependent (we will later come back to the important role of the lower bound of the integral). However, the mixing between growing and decaying (internal) modes produces a magnified signal in the folded configuration, e.g.~$k_1 \sim 2 k_2 \sim 2 k_3$, as the time integrals have one momentum with a flipped sign $\int^0\d\tau \, \tau^n e^{(-k_1+k_2+k_3)|c_s|\tau}$. Both signals are therefore present in the full bispectrum, and the relative amplitude and sign between the equilateral and folded limits depend on the micro-physical details of the tachyonic instability, e.g.~sign of $\rho_k$ and value of $\theta_k$. This magnification has been noticed in~\cite{Garcia-Saenz:2018vqf}, and we will get back to the value of this signal and perform a more detailed comparison later.

\vskip 4pt
From the exact bispectrum shapes produced by {\sf CosmoFlow} and given in Fig.~\ref{fig: hyperbolic shape examples} and Fig.~\ref{fig: Shape slices}, it follows that the folded-limit signal persists even in the case of a light entropic field with $\lambda=1$. In this regime, however, the effective description ceases to be valid, and only a genuine multi-field computation can capture this behavior. Moreover, the magnification in the folded limit is more pronounced for the interaction $(\partial_\mu\zeta)^2\sigma$ compared to the other operators. This behavior can be traced back to the additional momentum dependence, $\propto \bm{k}_2\cdot\bm{k}_3/k_1^2$ for the leading permutation, arising from the spatial-derivative contribution $(\partial_i\zeta)^2\sigma \subset (\partial_\mu\zeta)^2\sigma$, which dominates the scaling in the folded configuration.

\vskip 4pt
Finally, it is important to emphasize that the full bispectrum shape is given by a linear combination of all these contributions, with coefficients specified in Eq.~\eqref{eq: cubic interactions}. These Wilson coefficients depend sensitively on the underlying background dynamics. Since the different interactions exhibit distinct scalings in the folded limit, it is possible (although somewhat fine-tuned) for the strong signal in this limit to become negative relative to the equilateral configuration. This occurs when the Wilson coefficient of $(\partial_\mu \zeta)^2\sigma$ has the opposite sign and a larger magnitude than those of the other operators. As we will discuss in Sec.~\ref{sec: Concrete realization}, this situation is realized in angular inflation.

\subsection{Tachyonic resonance in mildly squeezed limit}
\label{subsec: tachyonic resonance}

\noindent
A striking feature in Fig.~\ref{fig: hyperbolic shape examples} and Fig.~\ref{fig: Shape slices} is the presence of a resonance in the mildly-squeezed kinematic configuration of the bispectrum shape, in case the entropic field is heavy.\footnote{Note that cosmological collider signals for heavy entropic fields are not visible in this figure as they appear in even softer kinematic configurations.} As $\lambda$ increases, corresponding to a stronger tachyonic instability, the amplitude of the resonance grows and its location is slightly deviated to softer kinematic configurations. 

\vskip 4pt
Note that, in the light entropic field case, this tachyonic resonance is completely dwarfed by the power-law scaling, that dominates soft kinematic configurations. Moreover, it is important to realise that this tachyonic resonance has a very different origin from those discussed in~\cite{Jazayeri:2022kjy, Wang:2022eop, Jazayeri:2023xcj}; in those works, the resonances arise from one or more reduced sound speeds (real $c_s \ll 1$) and are not correlated with a magnified folded limit, since no excited states are present.

\vskip 4pt
Instead, this resonant behavior can be qualitatively well understood within the single-field EFT framework introduced earlier and studied in e.g.~\cite{Garcia-Saenz:2018vqf}. As outlined there, this EFT cannot be trusted up to arbitrarily high energies or equivalently infinitely past times: the effective description of the curvature perturbation mode~\eqref{eq: EFT mode function} is valid only up to when the destabilized entropic field starts growing. This cut-off is set by the effective parameter $x$ that sets the regime of validity of the single-field EFT: 
 \begin{align}
     |c_s| k/(aH) \lesssim x \,,
 \end{align}
(at earlier times, both fields are decoupled in the Bunch-Davies vacuum). While this cut-off parametrises the same scale as our mass ratio $\lambda$, these two parameters are not necessarily identical and can be related via a non-trivial mapping.  In what follows, we will therefore be agnostic about $x$ and determine its value by fitting the bispectrum shape in e.g.~the isosceles folded configuration.

\vskip 4pt
With the EFT cut-off, the mixing between growing and decaying modes in~\eqref{eq: EFT mode function} therefore gives the contribution (among others) to the shape function (for the effective interaction $\dot{\zeta}^3$):
\begin{equation}
\label{eq: shape squeezed limit EFT}
    \begin{aligned}
        &S(k_1, k_2, k_3) \propto (k_1k_2k_3) \int_{-x/k_\star}^0\d\tau\, \tau^2 e^{(-k_1+k_2+k_3)\tau} \\
        &= \left(\frac{k_s}{k_\ell}\right)^2\left[1-\left(1+x\tfrac{k_\ell}{k_s} + \tfrac{1}{2}\left(x\tfrac{k_\ell}{k_s}\right)^2\right)e^{-x k_\ell/k_s}\right]\,,
    \end{aligned}
\end{equation}
where $k_\star \equiv \max(k_1, k_2, k_3)$, and we have chosen the permutation $k_2 = k_3 = k_s$ and $k_1 = k_\ell$ for the short and long modes, respectively. A few comments are in order:
\begin{itemize}
\item 
The exponential factor, originating from the EFT cutoff, acts as a regulator and ensures the shape vanishes in the squeezed limit $k_\ell / k_s \to 0$; in particular, the leading term in this soft limit is given by $\tfrac16 x^3 k_\ell / k_s$. 
\item 
In contrast, in mildly squeezed configurations, the above shape has a resonance whose location is found to be located at $k_\ell / k_s \simeq 1.45 / x$, in qualitative agreement with Fig.~\ref{fig: Shape slices}.\footnote{In the case of an asymmetric mass parametrization, the regime of validity of the EFT is given by $\tau >  -x/(|c_s| k_\star)$ instead. We have verified that the presence of the tachyonic resonance is independent of the mass parametrization~\eqref{eq: masses parametrization}, but its position is shifted towards the squeezed limit as the strength of the instability increases, see App.~\ref{app2} for details.}
\end{itemize}
The above example, focusing on the interaction~$\dot{\zeta}^3$, can be extended to any powers $n$ of the conformal time within the integral~\eqref{eq: shape squeezed limit EFT}, with e.g.~$n=1$ relevant for the cubic interaction $(\partial_i \zeta)^2\dot{\zeta}$. In this case, the shape function along isosceles kinematic configurations becomes
\begin{equation}
    S(k_1, k_2, k_3) \propto \left(\frac{k_\ell}{k_s}\right)^{-n}\left[1-e^{-x \tfrac{k_\ell}{k_s}}\left(e^{+x \tfrac{k_\ell}{k_s}}\right)_n\,\right]\,,
\end{equation}
where $(\cdot)_n$ denotes the truncation of the (exponential) function as a Taylor series at orders up to and including $n$. The ensuing discussion of squeezed and mildly squeezed limits is similar. 

\vskip 4pt
Therefore, we observe an interesting relation between the behaviour of the bispectrum shape in these mildly-squeezed and folded configurations, with the single-EFT of~\cite{Garcia-Saenz:2018vqf} offering a qualitative explanation of both features. This can serve as a consistency measurement to pinpoint inflationary models with a transient instability, as to our knowledge it cannot be reproduced by other effects.

\subsection{UV matching}

\begin{figure}[t!]
    \centering
    \includegraphics[width=7.5cm]{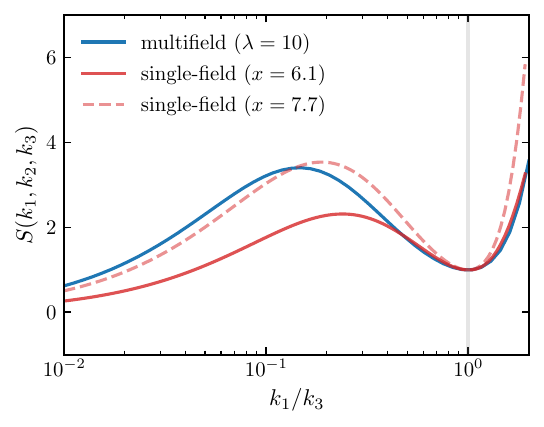}
    \caption{\it Dimensionless bispectrum shapes $S(k_1, k_2, k_3)$ in the isosceles configuration $k_2=k_3$ from the folded limit $k_1/k_3=2$ to the squeezed limit $k_1/k_3\to0$, fixing $\lambda=10$, from the exact multi-field calculation (\textcolor{pyblue}{\textit{blue}}) and as predicted by the single-field EFT (\textcolor{pyred}{\textit{red}}). The UV matching of the parameter $\lambda\to x$ is done by fitting the bispectrum amplitude at the isosceles folded configuration. For illustrative purposes, we have chosen the interaction $\dot{\zeta}\sigma^2$ which reduces to $\dot{\zeta}^3$ in the effective description. Notice that fitting the effective parameter $x$ at the tachyonic resonance peak would result in mismatch for its location and the amplitude in the folded contribution would be overestimated.}
    \label{fig: EFT mismatch}
\end{figure}

\noindent
The exact {\sf CosmoFlow} computations outlined in this paper allow for the previously inaccessible regime of light fields, with $\lambda$ of order one. However, it turns out that this approach also captures novel features in the heavy case, not captured by the analysis of~\cite{Garcia-Saenz:2018vqf}. This new multi-field feature ultimately can be traced back to a {\it kinematic-dependent nature} of the UV matching of the effective parameter $x$ which sets the regime of validity of the single-field EFT.
An example of this mismatch can be seen in Fig.~\ref{fig: EFT mismatch}. As we will see, this demonstrates that the regime of validity for the single-field EFT is kinematic dependent, and is restricted to kinematic configurations that do not feature any (even mild) hierarchy of scales.

\vskip 4pt
In order to compare predictions from the multi-field and single-field theories, we numerically compute the bispectrum shapes for the different interactions $(\partial_\mu \zeta)^2\sigma, \dot{\zeta}\sigma^2$ and $\sigma^3$, fixing a kinematic configuration, and varying the parameter $\lambda$. In the equilateral limit, the single-field EFT predictions are (without overall normalisation to one, and neglecting exponentially suppressed terms)
\begin{equation}
    S^{\dot{\zeta}^3}_{\rm EFT}(k, k, k) =  \frac{13}{6}\,, \quad S^{(\partial_i \zeta)^2\dot{\zeta}}_{\rm EFT}(k, k, k) = -\frac{5}{24} \,, \label{equilateral}
\end{equation}
whereas in the isosceles folded limit, these are given by
\begin{equation}
    \begin{aligned}
        S^{\dot{\zeta}^3}_{\rm EFT}(k/2, k/2, k) &=  \frac{1}{128} (39 + 4 x^3) \,, \\
        S^{(\partial_i \zeta)^2\dot{\zeta}}_{\rm EFT}(k/2, k/2, k) &= \frac{1}{128} (-39 + 12 x^2 + 4 x^3) \,.
    \end{aligned}
\end{equation}
Here, we have split the operator $(\partial_\mu\zeta)^2$ into $\dot{\zeta}^2$ and $(\partial_i\zeta)^2$ to have neat EFT expressions. Other kinematic configurations can be easily obtained from the shapes computed in~\cite{Garcia-Saenz:2018vqf}.

\begin{figure}[t!]
    \centering
    \includegraphics[width=7.5cm]{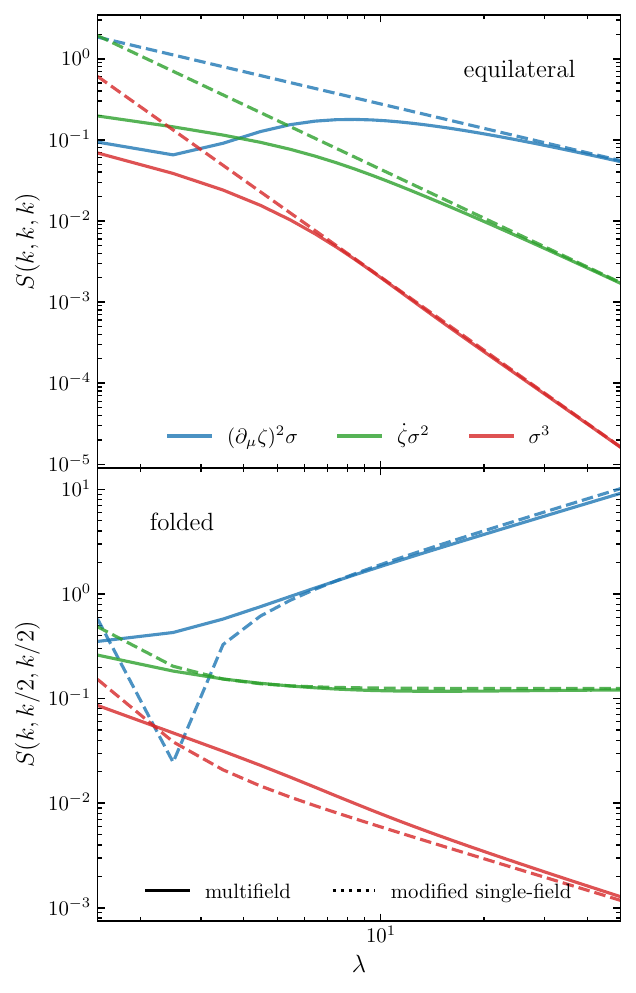}
    \caption{\it Dimensionless bispectrum shape functions in the equilateral $k_1=k_2=k_3=k$ ({\rm upper panel}) and isosceles folded $k_1=2k_2=2k_3=k$ ({\rm lower panel}) for all considered interactions, as predicted by the multi-field theory (solid line) and the modified single-field EFT (dashed line). Notice that we have rescaled the single-field EFT prediction as in~\eqref{eq: empirical scaling}, including the factor from integrating out the heavy field and the additional empirical factor $f(\{k_i\})/\lambda$ in the folded configuration only, which comes from a kinematic-dependent UV matching.}
    \label{fig: Comparison EFT}
\end{figure}

\vskip 4pt
To match the single-field predictions with the multi-field results, one needs to take into account the rescaling coming from integrating out the heavy field $\sigma$, as $\sigma=\sqrt{2}/(\lambda H) \, \dot{\zeta}_c$, which includes a $1/\lambda$ suppression. In the equilateral configuration, the single-field approach then reproduces the multi-field predictions:
\begin{equation}
\label{eq: empirical scaling-equil}
    \begin{aligned}
        {S}_{\rm multi}(k_1 = k_2 =k _3; \lambda) & \simeq \left(\frac{\sqrt{2}}{\lambda}\right)^n \\
        &\times S_{\rm single}(k_1 = k_2 = k_3; x) \,,
    \end{aligned}
\end{equation}
where $n$ matches the number of $\sigma$ fields in the considered interaction. We refer to the right-hand side of \eqref{eq: empirical scaling-equil} as the modified (or rescaled) single-field prediction. As shown in the upper panel of Fig.~\ref{fig: Comparison EFT}, this reproduces the multi-field result perfectly for $\lambda \geq 20$. Note that this does not require a precise matching between $x$ and $\lambda$, as the bispectrum shape in this limit is parameter-independent \eqref{equilateral}. Naturally, as $\lambda$ decreases, the single-field EFT prediction significantly deviates from the exact result, as expected.

\vskip 4pt
However, away from the equilateral limit, we find that matching between the single- and multi-field bispectrum shapes requires a kinematic-dependent matching (in addition to the aforementioned scaling with $\lambda$). In particular, along folded kinematic configurations (not necessarily isosceles), with $k_2 + k_3 \sim k_1$, we numerically find the following empirical scaling that is necessary to correct the single-field predictions:
\begin{equation}
\label{eq: empirical scaling}
    \begin{aligned}
        {S}_{\rm multi}(k_1 \sim k_2+k_3; \lambda) & \simeq \left(\frac{\sqrt{2}}{\lambda}\right)^n \frac{f(\{k_i\})}{\lambda} \\
        &\times S_{\rm single}(k_1 \sim  k_2 + k_3; x=\lambda) \,,
    \end{aligned}
\end{equation}
where the kinematic-dependent function $f$ is given by\footnote{We have checked this empirical kinematic dependent overall normalization also for regular folded but not isosceles $(k_1, k_2, k_3) = (4/5, 1/5, 1)k$, and folded albeit squeezed $(k_1, k_2, k_3) = (19/20, 1/20, 1)k$ configurations. In each case, we find a perfect match in the asymptotic large-$\lambda$ limit. Also, as $\lambda$ decreases, the results follow the expected trend, providing evidence that the higher-order terms in the large-$\lambda$ expansion are correctly reproduced.}
\begin{equation}
    f(k_1, k_2, k_3) = \frac{1}{2}\left(\frac{k_1}{k_2} + \frac{k_1}{k_3}\right)\,, \quad \text{(folded)} \,. \label{folded}
\end{equation} 
As one can see in Fig.~\ref{fig: Comparison EFT}, it matches the multi-field results well for larger values of $\lambda$. We stress that this factor has been introduced for empirical reasons; it would be very interesting (but challenging) to further understand its origin from a UV perspective.

\vskip 4pt
This comparison makes clear that the UV matching, giving rise to the relation between the UV parameter $\lambda$ and the effective parameter $x$, is kinematic dependent (with \eqref{folded} capturing folded configurations). As a result, it is not possible to fully accurately describe the bispectrum shape in all kinematic configurations simultaneously, with a single $x=x(\lambda)$, as illustrated in Fig.~\ref{fig: EFT mismatch}.

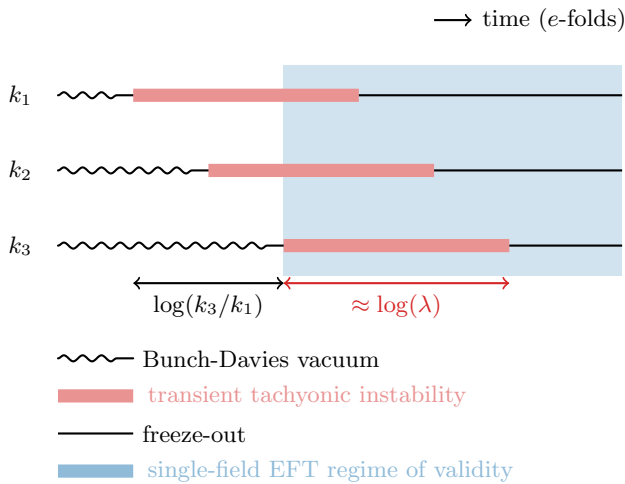
\begin{figure}[t!]
    \centering
    \begin{tikzpicture}[scale=1]
        \draw[->, black, thick, opacity=1] (1, 3) -- (1.5, 3);
		\node at (2.6, 3) {time ($e$-folds)};

        \draw[<->, pyred, thick, opacity=1] (-1, -0.5) -- (2, -0.5);
        \node at (0.5, -0.8) {\textcolor{pyred}{$\approx\log(\lambda)$}};

        \draw[<->, black, thick, opacity=1] (-3, -0.5) -- (-1, -0.5);
        \node at (-2, -0.8) {$\log(k_3/k_1)$};

        \fill[pyblue, fill opacity=0.2] (-1, -0.4) -- (3.5, -0.4) -- (3.5, 2.4) -- (-1, 2.4) -- cycle;
        
        \draw[BD] (-4, -1.5)  -- (-3, -1.5) node[right] {Bunch-Davies vacuum};

        \draw[pyred!50, line width=5pt] (-4, -2)  -- (-3, -2) node[right] {transient tachyonic instability};

        \draw[black, thick] (-4, -2.5)  -- (-3, -2.5) node[right] {freeze-out};

        \draw[pyblue!50, line width=5pt] (-4, -3)  -- (-3, -3) node[right] {single-field EFT regime of validity};
        
        \node at (-4.5, 2) {$k_1$};
        \draw[BD, shorten >=-2pt] (-4, 2)  -- (-3, 2);
        \draw[pyred!50, line width=5pt] (-3, 2) -- (0, 2);
        \draw[black, thick] (0, 2) -- (3.5, 2);

        \node at (-4.5, 1) {$k_2$};
        \draw[BD, shorten >=-2pt] (-4, 1)  -- (-2, 1);
        \draw[pyred!50, line width=5pt] (-2, 1) -- (1, 1);
        \draw[black, thick] (1, 1) -- (3.5, 1);

        \node at (-4.5, 0) {$k_3$};
        \draw[BD, shorten >=-2pt] (-4, 0)  -- (-1, 0);
        \draw[pyred!50, line width=5pt] (-1, 0) -- (2, 0);
        \draw[black, thick] (2, 0) -- (3.5, 0);
        
    \end{tikzpicture}
    \caption{\it Schematic illustration of the interplay between three modes $k_1<k_2<k_3$ as function of time (in $e$-folds). The leading single-field EFT contribution correlates the longest {\rm decaying} $k_1$ mode with two {\rm growing} modes $k_2$ and~$k_3$. The single-field EFT regime of validity $x$  is to be fitted to the data or determined by UV matching. The transient tachyonic instability described by this effective theory lasts $\approx \log(x)$ (which can differ from $\log(\lambda)$), and inevitably misses part of the evolution in case of different momenta.}
    \label{fig: understanding mismatch}
 \end{figure}

\vskip 4pt
This new kinematic-dependent scaling we find in folded configurations can be traced back to the regime of validity of the single-field EFT. Consider a general kinematic configuration coupling three different modes $k_1, k_2$ and $k_3$, which, without loss of generality, we order as $k_1 \leq k_2 \leq k_3$. The effective description is at most valid for $\tau\gtrsim - x/(|c_s|k_3)$ where $k_3$ is the {\it shortest} mode. This requirement ensures that the single-field EFT remains applicable throughout the entire time interval during which any given mode either undergoes a transient tachyonic instability (that roughly lasts $\sim \log(\lambda)$ $e$-folds) or has already frozen out. This generic configuration is schematically represented in Fig.~\ref{fig: understanding mismatch}. 

\vskip 4pt
For a generic kinematic configuration, the leading single-field bispectrum contribution comes from the interference of two growing and one decaying modes. The interference of all three (internal) growing modes actually is real and gives no contribution to the bispectrum, as shown in~\cite{Garcia-Saenz:2018vqf}.\footnote{The equilateral-like contribution is found by interfering internal growing modes with external decaying ones.} The dominant contribution thus comes from the permutation in which the decaying mode corresponds to the {\it longest} mode $k_1$---this mode undergoes less suppression from the tachyonic instability than the shorter modes within the time window over which the effective single-field EFT description is valid. On the other hand, the multi-field theory is valid throughout all times and takes into account the time interval during which the longest mode is exponentially decaying (which lasts $\Delta N \equiv N_{k_3}-N_{k_1} = \log(k_3/k_1)$ $e$-folds). This naturally results in a suppressed signal compared to the single-field prediction.

\vskip 4pt
From the above reasoning, it is clear that the exact equilateral configuration $k_1=k_2=k_3$ is well described by both single-field and multi-field theories, as all modes experience the tachyonic instability and then freeze out at the same moment. Instead, in the soft limit $k_3/k_1 \gg \lambda$ (with $k_2 \sim k_3$), and neglecting particle production, the long mode acts as background rescaling and the signal vanishes: this is the well-known single-field consistency condition. When the two characteristic scale ratios coincide $k_3/k_1 \approx \lambda$, the shape function exhibits the `tachyonic resonance' described in Sec.~\ref{subsec: tachyonic resonance}.

\subsection{Non-geodesic shape templates}

\noindent
The precise bispectrum shapes are in general complicated and require a numerical computation. However, the universal features of these `non-geodesic shapes' - applicable to slowly-varying backgrounds - motivate us to design simple templates for data analysis.

\vskip 4pt
Before doing so, it is important to recall that the full observed shape is obtained by summing over all contributions from the cubic interactions in~\eqref{eq: cubic interactions}. The relative weight of each individual shapes depends on the precise values and signs of the relevant Wilson coefficients. Moreover, as we have seen previously, $(\partial_\mu \zeta)^2\sigma$ leads to a larger magnification in the folded configuration, while $\sigma^3$ produces a stronger tachyonic resonance in mildly soft configurations compared to, for example, $\dot{\zeta}\sigma^2$. As a result of the interplay between different interactions, the folded limit can be large and positive or negative compared to the equilateral configuration. This important feature arises because the shapes produced by the three independent cubic operators~\eqref{eq: cubic interactions} are not identical, and we will introduce templates for both possibilities in the folded limit.

\begin{figure}[t!]
    \centering
    \begin{subfigure}{0.4\textwidth}
        \centering
        \includegraphics[width=0.9\textwidth]{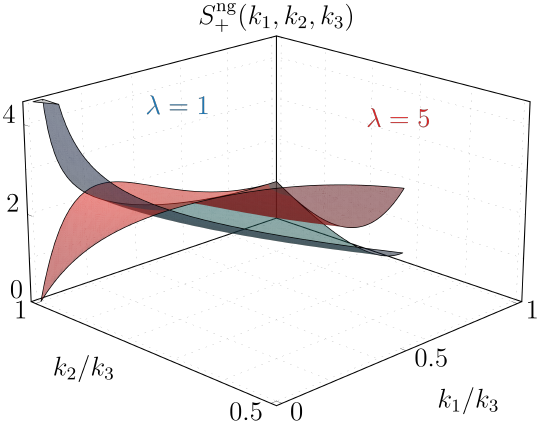}
    \end{subfigure}\hfill
    \begin{subfigure}{0.4\textwidth}
        \centering
        \includegraphics[width=0.9\textwidth]{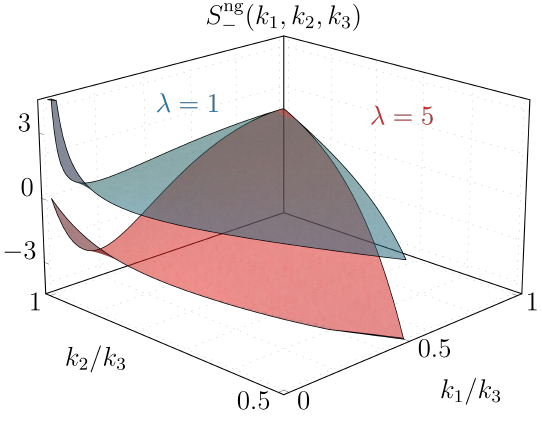}
    \end{subfigure}\hfill
    \caption{\it Non-geodesic shapes in all kinematic configurations for a large positive folded limit $S^{\rm ng}_+$ ({\rm top panel}) and for a large negative folded limit $S^{\rm ng}_-$ ({\rm bottom panel}), both for $\lambda=1$ ({\rm \textcolor{pyblue}{blue}}), corresponding to a light entropic field, and $\lambda=5$ ({\rm \textcolor{pyred}{red}}), which corresponds to stronger tachyonic instability and therefore to a heavy entropic field. Note the sign difference between the equilateral and folded limits for $S^{\rm ng}_-$. We have normalized the shapes to be unity in the equilateral configuration.}
    \label{fig: hyperbolic shape templates}
\end{figure}

\vskip 4pt
For the light entropic field case with $|m_\sigma|/H \leq 3/2$, corresponding to the case of a mild instability, we propose the following non-geodesic shape template:
\begin{equation}
\label{eq: light hyperbolic template}
    \begin{aligned}
        S^{\rm ng}_\pm(k_1, k_2, k_3) &= \tfrac{3\pm1}{2}\, S^{\rm flat}(k_1, k_2, k_3) \\
        &+ S^{\rm light}(k_1, k_2, k_3)\,, \quad (\lambda\leq 3/2)\,,
    \end{aligned}
\end{equation}
where $S^{\rm flat} \equiv -S^{\rm eq}+1$ is the flattened template~\cite{Meerburg:2009ys}, defined in terms of the standard equilateral shape~\cite{Creminelli:2005hu}
\begin{equation}
    \begin{aligned}
        S^{\rm eq}(k_1, k_2, k_3) &= \left(\frac{k_1}{k_2} + {\rm 5 \,perms}\right) \\
        &- \left(\frac{k_1^2}{k_2 k_3} + {\rm 2 \,perms}\right) - 2\,,
    \end{aligned}
\end{equation}
and $S^{\rm light}$ is a shape proposed in~\cite{Chen:2009zp} describing the soft-limit power-law scaling that depends on the dimensionless parameter $\lambda$ through $\nu_\lambda \equiv \sqrt{9/4-\lambda^2}$:
\begin{equation}
    S^{\rm light}(k_1, k_2, k_3) = 3\left(\frac{3k_1k_2k_3}{k_t^3}\right)^{\frac{1}{2}} \frac{N_{\nu_\lambda}\left(8k_1k_2k_3/k_t^3\right)}{N_{\nu_\lambda}(8/27)}\,,
\end{equation}
where $k_t \equiv k_1+k_2+k_3$ and $N_{\nu_\lambda}$ is the Neumann function. The $\pm$ parameter in~\eqref{eq: light hyperbolic template} dictates whether the folded limit has a strong positive ($+$) or negative ($-$) signal. Fig.~\ref{fig: hyperbolic shape templates} shows the light non-geodesic shape template $S^{\rm ng}_\pm$ for $\lambda=1$ in blue. The case ($+$) reproduces very well the shape $S^{(\partial_\mu \zeta)^2\sigma}$ for $\lambda=1$ depicted in Fig.~\ref{fig: hyperbolic shape examples}, which can also be seen in Fig.~\ref{fig: Shape slices}. 

\vskip 4pt
In the case of a heavy entropic field with $|m_\sigma|/H \geq 3/2$, corresponding to a strong instability, we propose the following template
\begin{equation}
\label{eq: heavy hyperbolic template}
    \begin{aligned}
        &S_\pm^{\rm ng}(k_1, k_2, k_3) = -f_\pm(x) \, \frac{k_1k_2k_3}{k_t^3} \\
        &+ \frac{k_1k_2k_3}{\tilde{k}_1^3}\left[1 - \left(1+\lambda\tfrac{\tilde{k}_1}{k_\star}+\tfrac{1}{2}\left(\lambda\tfrac{\tilde{k}_1}{k_\star}\right)^2\right)e^{-\lambda\tilde{k}_1/k_\star}\right] \\
        &+e^{-\pi\mu_\lambda} \frac{k_1k_2}{k_{12}^2}\left(\frac{k_3}{k_{12}}\right)^{1/2}\cos\left[\mu_\lambda \log\left(\tfrac{k_3}{2k_{12}}\right)\right] \\
        &+ {\rm 2 \,perms}\,, \quad (\lambda\geq 3/2)\,,
    \end{aligned}
\end{equation}
with $f_\pm \equiv 1+3\lambda(1\mp 1)$, $\tilde{k}_1 \equiv -k_1+k_2+k_3$, $k_{12} \equiv k_1+k_2$ and $\mu_\lambda \equiv \sqrt{\lambda^2-9/4}$. As exact analytical computations are out of reach, this template was found empirically, partly based on the single-effective description introduced earlier. Notice that the first and second lines come from the shape generated by $\dot{\zeta}^3$, computed in~\cite{Garcia-Saenz:2018vqf}, that we have slightly modified to also account for the possibility of a negative signal in the folded-limit. This template is shown in Fig.~\ref{fig: hyperbolic shape templates} for $\lambda=5$, again showing agreement with the exact shapes in Fig.~\ref{fig: hyperbolic shape examples} and Fig.~\ref{fig: Shape slices}. The distinctive observational signature of this scenario is the simultaneous presence of a tachyonic resonance in the mildly-squeezed configuration and a magnification of the folded limit.

\begin{figure}[t!]
    \centering
    \includegraphics[width=8.5cm]{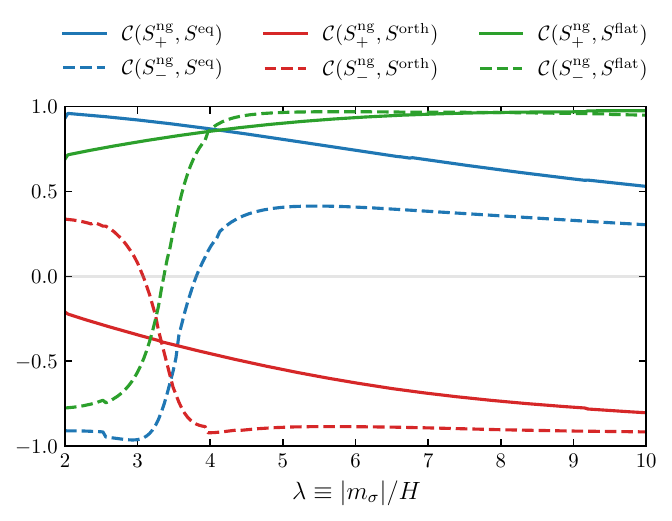}
    \caption{\it Correlations between the non-geodesic shape templates $S^{\rm ng}_\pm$ and the standard equilateral $S^{\rm eq}$, orthogonal $S^{\rm orth}$ and flattened $S^{\rm flat}$ shape templates as function of $\lambda$, for the heavy entropic field case.}
    \label{fig: shape cosines}
\end{figure}

\vskip 4pt
It will be useful to perform a quantitative comparison between the new non-geodesic shapes and the standard shape templates used in data analysis, namely equilateral $S^{\rm eq}$, flattened $S^{\rm flat}$, and orthogonal $S^{\rm orth}\equiv 3S^{\rm eq}-2$~\cite{Senatore:2009gt}. We compute their shape correlations $\C(S^{\rm ng}_\pm, S^b)$, as defined in~\cite{Babich:2004gb, Fergusson:2008ra}, as function of $\lambda$, where two shapes are strongly correlated if $|\C|\approx 1$.

\vskip 4pt
The shape correlation for the heavy entropic field case is displayed in Fig.~\ref{fig: shape cosines}.\footnote{For the light entropic field case, the cosine is ill-defined as the momentum integral scales as $\sim \int_0^1\d t\,t^{1-2\nu_\lambda}$, which converges only for $\lambda>\sqrt{2}$.} Both the positive and negative signals in the folded-limit non-geodesic shapes $S^{\rm ng}_\pm$ are weakly correlated with the equilateral template as $\lambda$ increases. This behavior is natural, since a stronger tachyonic instability amplifies the folded limit compared to the equilateral one. At large $\lambda$, the non-geodesic shapes become highly correlated with the flattened template. Obviously, $S^{\rm ng}_-$ also develops a strong correlation with the orthogonal shape in this regime. This is consistent with the fact that the orthogonal template features equilateral and folded contributions of opposite signs. Conversely, as $\lambda$ decreases, the correlation between $S^{\rm ng}_\pm$ and the equilateral template increases, reflecting the reduced signal in the folded limit. Finally, we note that the correlation of $S^{\rm ng}_-$ systematically changes sign in the range $3 \lesssim \lambda \lesssim 4$, implying the existence of a value of $\lambda$ at which $S^{\rm ng}_-$ is exactly ``orthogonal'' to each of the standard templates individually.

\section{Concrete case: angular inflation}
\label{sec: Concrete realization}

\noindent
We now turn to concrete realisations of background models that realise an entropic transient instability. There is a wide range of slow-roll fast-turn attractors that allow for this behaviour, see e.g.~\cite{Brown:2017osf, Mizuno:2017idt, Christodoulidis:2018qdw, Bjorkmo:2019aev, Fumagalli:2019noh, Garcia-Saenz:2018ifx, Achucarro:2012yr, Aragam:2019omo, Achucarro:2019mea, Achucarro:2019pux}. Note that the fast turn induces the trajectory to deviate from the potential valley that slow-roll slow-turn models follow. This can make them difficult to recognise; see e.g.~\cite{Aragam:2020uqi, Anguelova:2024akm, Wolters:2024vzk} for recent developments in searching for such trajectories. Before delving into a concrete example, let us first show that such models are generically found in negatively-curved field spaces by deriving universal bounds on background parameters to realize the destabilization of entropic fluctuations.

\subsection{Universal bound on background dynamics}
\label{sec: bound on background dynamics}
\noindent
A negative bare entropic mass arises from an interplay between the stabilizing potential, the turn rate and the field-space curvature, with no {\it a priori} hierarchy among these contributions. In what follows, we derive a simple yet broadly applicable bound on these background quantities that determines when entropic fluctuations undergo a transient tachyonic instability. 

\vskip 4pt
We will concentrate on the simplified case of a maximally symmetric two-dimensional field space (see~\cite{Aragam:2021scu} for more general situations). Employing polar coordinates $(\rho, \theta)$, the field-space metric reads 
 \begin{align}
&      \d s^2 = \d\rho^2 + S_K^2(\rho)\d\theta^2 \,, \notag  \\  
 &     S_K(\rho) = \{\M \sinh(\rho/\M), \rho, \M \sin(\rho/\M)\} \,,
       \end{align}
for negative ($K=-1$), flat ($K=0$) or positive ($K=+1$) curvature $R_{\rm fs} = 2K/\M^2$, respectively. We assume a general potential $V(\rho, \theta)$ and consider that the inflationary trajectory is stabilized around $\rho\approx \rho_0$ (with $\dot{\rho} \sim \ddot{\rho}\sim 0$) and instead performs slow roll along the angular direction (with $\ddot{\theta} \sim 0$). Under these conditions, the Hubble parameter and its variation follow from the Einstein equations and read
\begin{equation}
  3\Mpl^2H^2 = V \,, \quad  2\Mpl^2H^2 \epsilon = S_K^2(\rho)\dot{\theta}^2 \,.
\end{equation}
Moreover, the scalar field equations give
\begin{equation}
    S_K(\rho)S_K'(\rho)\dot{\theta}^2 = V_{,\rho}\,, \quad 3H\dot{\theta} + \frac{V_{,\theta}}{S_K^2(\rho)}=0 \,,
\end{equation}
where $S_K'(\rho) \equiv \tfrac{\d}{\d\rho}S_K(\rho)$. Combining these equations to eliminate $\dot{\theta}$ yields the trajectory equation $\rho = \rho(\theta)$.\footnote{In principle, one should also check that given a specific potential $V$, these equations are consistent with a slow-roll dynamics along the angular direction and an almost constant radius. This is model dependent and should be done on a case-by-case basis. More specifically, in order to consistently ignore $\dot{\rho}$, we need $\dot{\rho} \ll S_K \dot{\theta}$. This then implies $\d\rho/\d\theta \ll S_K$, which gives a model dependent condition.} To see whether such trajectories display a transient instability, we turn to the quantities that set the EFT coefficients. First of all, the turn rate is 
\begin{align}
    \eta_\perp  = \sqrt{2\epsilon} \, f_K(\rho) \,, \quad f_K(\rho) \equiv \Mpl S_K'(\rho)/S_K(\rho) \,,
\end{align}
and this is set by the first slow-roll parameter magnified by a geometric factor. This allows one to rewrite the bare entropic mass as
\begin{equation}
    \begin{aligned}
        \frac{m_\sigma^2}{H^2} &= \frac{V_{,\rho\rho}}{H^2} - 2\epsilon \left[f_K^2(\rho)- K\left(\frac{\Mpl}{\M}\right)^2\right]\,, 
    \end{aligned}
\end{equation}
We further assume that the stabilizing effect from the potential in the orthogonal direction is dictated by 
 \begin{align}
     V(\rho) = \tfrac{1}{2}M^2\rho^2 \,,
 \end{align}
and thus introduces a second scale $M$. One can identify two limiting regimes: (i) the negligible curvature limit $\rho_0/\M\ll1$, for which we have $f_K(\rho_0)\approx \Mpl/\rho_0$, and (ii) the large (and negative) curvature limit $\rho_0/\M \gg1$, so that $f_K(\rho_0) \approx \Mpl/\M$. Requiring a negative bare entropic mass yields the following bound
\begin{equation}
\left.
    \begin{array}{ll}
    \rho/H & \text{(small curvature)} \\
    \M/H  & \text{(large curvature)}
    \end{array}
\right\} \leq \sqrt{2\epsilon} \, \frac{\Mpl}{M} = \O(10^3)\frac{H}{M}\,,
\end{equation}
where we have taken the characteristic value $\epsilon \sim 10^{-2}$ and have considered that the Hubble scale can be as high as $10^{14}$ GeV. 

\vskip 4pt
We can observe that in the small curvature limit, entropic fluctuations become unstable when the trajectory radius is sufficiently small, such that the potential is no longer able to stabilize motion along the valley. On the contrary, when the field-space curvature dominates, it should be parametrically smaller than the potential stabilizing scale. In concrete UV completions, the field-space geometry is often hyperbolic, with its curvature set by the scale $\M$ typically of order the Kaluza–Klein scale, $\M \sim 10^{-2}\Mpl$ (see~\cite{Baumann:2014nda} for a review). This implies $M/H \lesssim 15$, consistent with most examples in the literature. 

\begin{figure}[t!]
    \centering
    \includegraphics[width=8.5cm]{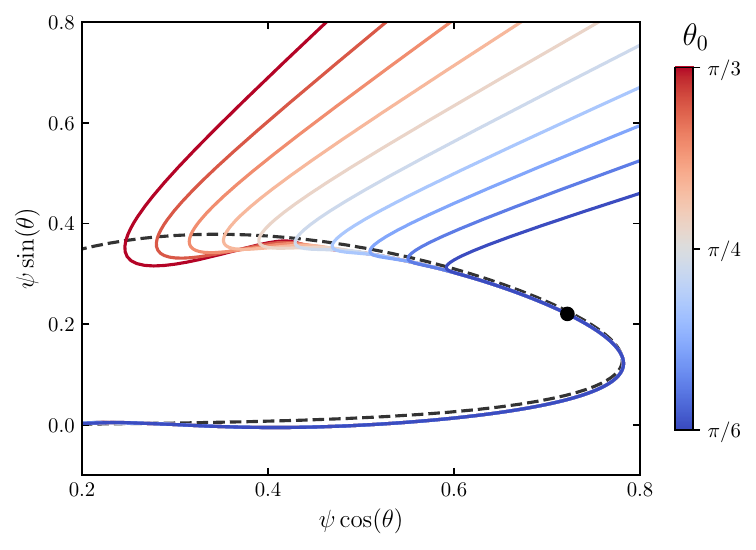}
    \caption{\it Background trajectories for $R_m=31$ and $\alpha=10^{-2}$ for an initial radius $\rho_0=0.9999$ for various initial angles $\theta_0$. We represent the approximate angular attractor~\eqref{eq: angular attractor} with the black dashed line. The black marker corresponds to the CMB window, $55$ $e$-folds before the end of inflation.}
    \label{fig: angular inflation background}
\end{figure}

\subsection{Background}

\noindent
In the rest of this section, we study a concrete background realization of fast-turn attractors: angular inflation~\cite{Christodoulidis:2018qdw} (see \cite{Iarygina:2023msy} for the first analysis of its bispectrum). 
We consider a two-dimensional field space, with fields $\phi^1 \equiv \phi$ and $\phi^2 \equiv \chi$, with the following field-space metric
\begin{equation}
    G_{IJ} = \frac{6\alpha}{(1 - \phi^2 - \chi^2)^2}\, \delta_{IJ}\,.\label{G_IJ_alpha}
\end{equation}
This describes a hyperbolic space (Poincaré disk) with constant negative curvature given by $\Mpl^2R_{\rm fs} = -\tfrac{4}{3\alpha}$. In the limit $\alpha \to \infty$, the field space becomes effectively flat. The potential is taken to be\footnote{One can of course redefine the fields to absorb the $\alpha$ parameter so that it does not appear in the potential. This choice does not affect observables.}
\begin{align}
    V(\phi, \chi) = \tfrac{\alpha}{2}\left(m_\phi^2 \phi^2 + m_\chi^2 \chi^2\right) \,,\label{V_alpha}
\end{align}
and we define $R_m \equiv m_\chi^2/m_\phi^2>1$ to be the mass ratio. It is natural to introduce polar coordinates $\phi\equiv \rho\cos(\theta)$ and $\chi\equiv \rho\sin(\theta)$ and further rescale the radial direction by defining $\psi \equiv \sqrt{6\alpha} \, \tanh^{-1}(\rho)$, since the field-space manifold is spherically symmetric. 

\vskip 4pt
As the mass of the light field $m_\phi$ dictates the overall amplitude of the primordial power spectrum, fixed by observations, this model has two free parameters: $R_m$ and $\alpha$. Typically when $\alpha \ll 1$ and $R_m \gtrsim 1$, the background trajectory experiences a phase of so-called angular inflation during which $\dot{\rho}\approx 0$. We depict this angular attractor in Fig.~\ref{fig: angular inflation background}. This arises from the competition between two effects: the hyperbolic target space pushes both fields to the boundary of the Poincaré disk, whereas the gradient of the potential attracts them to the origin. This regime can be solved analytically, which, in the regime $\alpha \ll 1$, yields~\cite{Christodoulidis:2018qdw}
\begin{equation}
\label{eq: angular attractor}
    1-\rho^2(\theta) \approx \frac{9\alpha(\cot\theta+R_m \tan\theta)^2}{2(R_m-1)^2} \,,
\end{equation}
and $\epsilon \approx \tfrac{3}{2}(1-\rho^2)$. The number of $e$-folds during the angular phase is found to be $N\approx \tfrac{R_m}{27\alpha}+\ldots$, valid for $R_m \gg1$, where higher-order logarithmic corrections have been computed in~\cite{Christodoulidis:2018qdw}. Note that the angular phase is preceded by an effective single-field inflationary trajectory in the radial direction. 

\begin{figure}[t!]
    \centering
    \includegraphics[width=8.5cm]{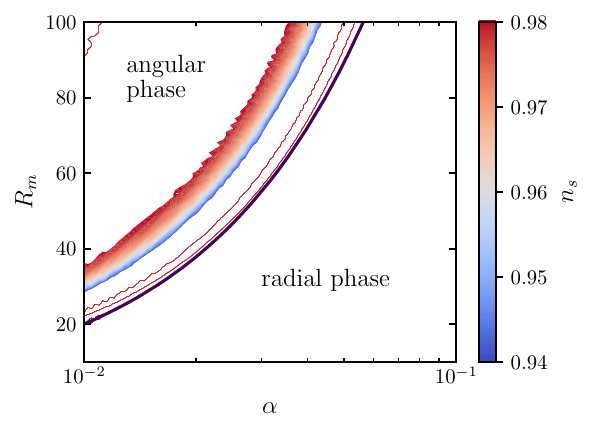}
    \caption{\it Spectral tilt $n_s$ in the two-dimensional parameter space $(\alpha, R_m)$. Below the solid dark line, the angular phase lasts less than $60$ $e$-folds and the CMB window is located in the radial phase. Note that this radial inflationary phase, for which we have not computed $n_s$, is not excluded by observations as it is equivalent to single-field slow-roll inflation.}
    \label{fig: angular phase diagram}
\end{figure}

\vskip 4pt
In Fig.~\ref{fig: angular phase diagram}, we show the region in the two-dimensional parameter space $(\alpha, R_m)$ that supports at least $60$ $e$-folds in the angular phase before the end of inflation. In this region, we have computed the scalar spectral tilt $n_s\equiv \partial \P_\zeta/\partial \log k+1$ using five points around the pivot scale $k_\star$ that exits the horizon $55$ $e$-folds before the end of inflation. One can observe a narrow band region that supports angular inflation still compatible with Planck results $n_s=0.9649\pm0.0042$~\cite{Planck:2018jri} and the recent ACT update $n_s=0.974 \pm 0.003$~\cite{ACT:2025fju}, when combined with DESI~\cite{DESI:2024mwx} and Planck data. We verified that small variations in the initial conditions, $r_0$ and $\theta_0$, do not significantly affect the results.

\subsection{Bispectrum shape}
\label{subsec: bispectrum shape}

\noindent
In the angular phase with approximate slow-roll solution~\eqref{eq: angular attractor}, projecting the first derivative of the potential onto the normal direction gives the turn rate 
\begin{equation}
    \eta_\perp^2 = \frac{4\epsilon}{3\alpha} \,.
\end{equation}
Note that for a field-space curvature of the order of slow-roll $\alpha \sim \epsilon$, order one turn rate $\eta_\perp =\O(1)$ is realized. The bare and effective entropic masses are found to be
\begin{equation}
    m_\sigma^2 = -3\eta_\perp^2H^2 \,, \quad m_{\sigma, \text{eff}}^2 = \eta_\perp^2 H^2\,.
\end{equation}
The entropic fluctuation therefore experiences a transient tachyonic instability before decaying on super-horizon scales. Notice the asymmetry between $m_\sigma^2$ and $m_{\sigma, \text{eff}}^2$ compared to the parametrization~\eqref{eq: masses parametrization}. However, we recall that we find no significant changes when independently varying both masses. 

\begin{figure}[t!]
    \begin{subfigure}{0.4\textwidth}
        \centering
        \includegraphics[width=0.9\textwidth]{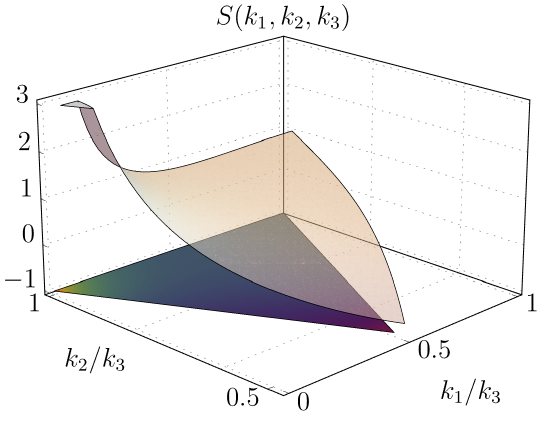}
    \end{subfigure}
    \caption{\it Bispectrum shape in all physical kinematic configurations for angular inflation with $R_m=31$ and $\alpha=10^{-2}$. We have chosen $r_0=0.999$ and $\theta_0 = \pi/4$ as initial conditions.}
    \label{fig: angular inflation shape}
\end{figure}

\vskip 4pt
In Fig.~\ref{fig: angular inflation shape}, we show the full bispectrum shape for $(\alpha, R_m) = (10^{-2}, 31)$, that gives a spectral tilt compatible with Planck data. The numerical computation was performed using \textsf{PyTransport}~\cite{Dias:2016rjq, Mulryne:2016mzv, Ronayne:2017qzn}. The overall non-Gaussian signal in the equilateral configuration is found to be $\fnl(k, k, k) = -0.2$. The shape displays the characteristic features discussed previously. As also found in \cite{Iarygina:2023msy}, the equilateral and folded kinematic configurations have opposite signs, which comes from a linear combination of the shapes $S^{(\partial_\mu \zeta)^2\sigma}$, $S^{\dot{\zeta}\sigma^2}$ and $S^{\sigma^3}$ computed previously with coefficients dictated by the background evolution. Notably, the shape exhibits an enhanced squeezed limit which shows that the (effective) entropic fluctuation is light. This is opposed to e.g.~sidetracked inflationary models~\cite{Garcia-Saenz:2018ifx, Fumagalli:2019noh} in which the entropic fluctuations are heavy. This shape is well captured by the template $S^{\rm ng}_-$ for $\lambda\approx 1$, shown in Fig.~\ref{fig: Shape slices}. We notice that the template is not perfect though as a linear combination of non-linear sigma model cubic operators cannot result in a local minimum in mildly-soft configurations, but only in an inflection point.

\subsection{SUGRA embedding}

\noindent
Finally, we briefly discuss a supergravity embedding of the multi-field $\alpha$-attractor system~\eqref{G_IJ_alpha} and~\eqref{V_alpha}.

\vskip 4pt
Following~\cite{Yamada:2018nsk,Linde:2018hmx}, we consider $\alpha$-attractors endowed with an approximate global U(1) symmetry.
Introducing the inflaton multiplet $\Phi$ and, for simplicity, a nilpotent multiplet $S$ satisfying $S^2=0$,\footnote{Alternatively, one
may introduce another chiral multiplet $S$ (the so-called stabilizer) and consider a situation in which it is decoupled via a strong stabilization mechanism at $S=0$.} we adopt the following Kähler potential and superpotential:
\begin{align}
K=-3 \alpha \log (1-|\Phi|^2)+G_{S \bar{S}} |S|^2\,,\quad W  =c+S\,,
\end{align}
where
\begin{align}
&G_{S \bar{S}}\equiv \left[(1-|\Phi|^2)^{3 \alpha} V+3\left|c\right|^2(1-\alpha|\Phi|^2)\right]^{-1}\,,\\
&V\equiv \frac{\alpha}{4}(m^2_\phi+m^2_\chi)|\Phi|^2+\frac{\alpha}{8}(m^2_\phi-m^2_\chi)(\Phi^2+\bar{\Phi}^2)\,,
\end{align}
and $c$ is a constant. The term proportional to $m^2_\phi-m^2_\chi$ explicitly breaks the U(1) symmetry $\Phi \rightarrow \Phi e^{i\theta}$.
Under the projection $S=0$, this construction reproduces the system~\eqref{G_IJ_alpha} and~\eqref{V_alpha}, with the identifications
$\phi\equiv {\rm Re}\,\Phi$ and $\chi \equiv {\rm Im}\,\Phi$. 

\vskip 4pt
Owing to the decoupling of the $S$ field, no additional features arise in the scalar sector, and the resulting dynamics coincide with those of the corresponding non-supersymmetric system. A genuine distinction from the non-supersymmetric theory would require a detailed analysis of the fermionic sector, and possibly of the reheating dynamics, which is beyond the scope of the present work.

\section{Conclusions}
\label{sec: conclusions}

\noindent
The primary goal of this work was to investigate the impact of non-geodesic background motion in internal field space on the inflationary bispectrum. Our analysis was carried out entirely at the level of fluctuations, without assuming any specific background model. This was possible because, commonly in negatively curved field spaces and in the absence of stabilizing potential effects along the entropic direction, entropic fluctuations generically undergo a transient tachyonic instability before horizon crossing. Indeed, a large turning rate drives the bare entropic mass negative, and this effect is further amplified by hyperbolic geometries. The subsequent sourcing of exponentially enhanced entropic fluctuations into the curvature perturbation sector imprints distinctive signatures in the bispectrum.

\vskip 4pt
Focusing on perfectly scale-invariant scenarios by neglecting variations of the Hubble scale, we parameterized the negative bare and positive effective entropic masses in terms of the turning rate, and computed the bispectrum shapes arising from all cubic operators unsuppressed by slow roll, giving order-one non-Gaussian signals. The resulting shapes reveal characteristic features tied both to the entropic mass and to the field-space geometry. In particular, in the squeezed limit, the bispectrum exhibits the non-analytic scaling, the so-called cosmological collider signal, with the scaling exponent determined by the entropic effective mass on super-horizon scales. Light entropic fields yield power-law enhancements, whereas heavy entropic fields produce distinctive oscillatory patterns.

\vskip 4pt
Our analysis allowed us to identify previously known magnified folded configurations, which can be positive or negative relative to the equilateral limit, which come from interference between growing and decaying curvature modes, triggered by the excited states generated during the tachyonic instability. Furthermore, we characterize a distinctive resonance in mildly soft kinematic configurations, whose position is directly linked to the strength of the instability. The combined presence of this resonance together with an amplified folded signal would provide a clean observational signature of geometrically induced tachyonic instabilities, distinguishing it from non-Bunch-Davies initial states that can mimic only the latter. 

\vskip 4pt
Our results build on earlier studies restricted to heavy entropic fields, where an effective single-field description applies. We find that the UV matching of the effective parameters of a single-field description intrinsically depends on kinematics for folded configurations. This implies that there is no UV matching that would accurately reproduce the bispectrum shape across all kinematic configurations simultaneously.
Therefore, a genuine multi-field calculation is needed to fully capture the effects of this transient instability. Moreover, our analysis extends this bispectrum also to the case of light entropic fields, where we find a similar magnification in the folded limit.

\vskip 4pt
For data analysis, we introduced a `non-geodesic' bispectrum template that faithfully captures all these distinctive features. To illustrate the robustness of our predictions, we then considered a concrete background model where the trajectory undergoes a phase of angular inflation in negatively curved field space. Scanning the relevant two-dimensional parameter space (field mass ratio and field-space curvature), we computed the scalar spectral tilt and identified a region compatible with observations. We then evaluated the bispectrum shape numerically, confirming that all the predicted features are indeed realized. Eventually, we exhibited a supersymmetric embedding of this angular inflation setup, providing a UV completion of our scenario.

\vskip 4pt
In summary, our model-independent analysis, reinforced by a concrete UV realization, highlights robust observational signals of slowly-varying non-geodesic motion in field space, naturally arising in hyperbolic geometries, that can be tested with current and upcoming cosmological surveys.

\vskip 4pt
A possible but non-trivial direction for future work is to extend the present analysis to multi-field systems with $N>2$, which are well-motivated in various contexts, including string-inspired scenarios such as N-flation~\cite{Dimopoulos:2005ac,Dias:2018pgj}. In such setups, richer phenomena are expected to emerge, including self-interactions of entropy modes induced by the presence of torsion~\cite{Christodoulidis:2023eiw,Christodoulidis:2022vww}, more intricate target-space geometries~\cite{Pinol:2020kvw}, and the superposition of cosmological collider signals~\cite{Aoki:2020zbj,Pinol:2021aun}.
A detailed investigation of these features is left for future work.

\vskip 4pt
\noindent {\bf Acknowledgements.---}
It is our pleasure to thank Ana Ach\'ucarro, Perseas Christodoulidis, Sebastian Garcia-Saenz, Oksana Iarygina, Sadra Jazayeri, Marieke Postma, Margherita Putti and S\'ebastien Renaux-Petel for stimulating discussions. We especially thank Sebastian Garcia-Saenz and S\'ebastien Renaux-Petel for useful comments on the draft.

\vskip 4pt
This work was made possible by the program CoBALt held at the Institut Pascal at Université Paris-Saclay with the support of the program ``Investissements d’avenir'' ANR-11-IDEX-0003-01. S.A. is supported by the Japan Science and Technology Agency (JST) 
as part of Adopting Sustainable Partnerships for Innovative Research Ecosystem (ASPIRE), 
grant JPMJAP2318, and by JSPS KAKENHI Grant Number 26K07078. D.W. is funded by the European Research Council under the European Union’s Horizon 2020 research and innovation programme (grant agreement No 758792, Starting Grant project GEODESI). D.W. is further supported by the European Union (ERC, \raisebox{-2pt}{\includegraphics[height=1\baselineskip]{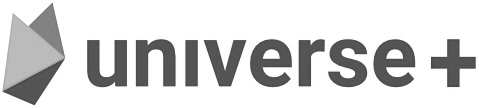}}, 101118787).

\vskip 4pt
This article is distributed under the Creative Commons Attribution International Licence (\href{https://creativecommons.org/licenses/by/4.0/}{CC-BY 4.0}). Views and opinions expressed are however those of the author(s) only and do not necessarily reflect those of the European Union or the European Research Council Executive Agency. Neither the European Union nor the granting authority can be held responsible for them.

\appendix
\section{Consistency of super-horizon \& single-field EFT limits}
\label{app}

\noindent
In this appendix, we address the consistency of the heavy-$\sigma$ single-field EFT limit ($|m_\sigma^2| \gg H^2$) and the super-horizon limit ($k/a \ll H$). These limits correspond to $\sigma = 2\eta_\perp H/m_\sigma^2 \, \dot{\zeta}_c$ and $\sigma = 1/(2\eta_\perp H)\dot{\zeta}_c$, respectively, which can be derived from the equations of motion of the quadratic action~\eqref{quadratic}. With our choice~\eqref{eq: masses parametrization}, these substitutions read $\sigma = \sqrt{2}/(\lambda H) \, \dot{\zeta}_c$ and $\sigma = 1/(\sqrt{2}\lambda H)\dot{\zeta}_c$, respectively, which naively seem not to be consistent. 

\vskip 4pt
Let us start by assuming the super-horizon limit $\sigma = 1/(2\eta_\perp H)\dot{\zeta}_c$. Inspecting the equation of motion for $\sigma$: $\ddot{\sigma}+3H\dot{\sigma} + m_{\sigma, {\rm eff}}^2\sigma=0$, the large-mass limit allows us to neglect the first two terms, yielding $m_{\sigma, {\rm eff}}^2\sigma=0$. The solutions are either: (i) $\sigma=0$ which gives $\dot{\zeta}_c=0$, i.e.~$\zeta_c$ is conserved, or (ii) $m_{\sigma, {\rm eff}}^2 = 0$, i.e.~$m_\sigma^2 = -4\eta_\perp^2H^2$.\footnote{This regime is referred to in the literature as the ultralight regime, see e.g.~\cite{Achucarro:2016fby}.}

\vskip 4pt
Conversely, let us assume instead the large-mass limit, $\sigma = 2\eta_\perp H \dot{\zeta}_c/m_{\sigma}^2$. The equation of motion for $\zeta_c$ is given by
\begin{equation}
    \frac{1}{a^3}\frac{\d}{\d t} \left(a^3 \dot{\zeta}_c \frac{m^2_{\sigma, {\rm eff}}}{m_\sigma^2}\right) + \frac{k^2}{a^2}\zeta_c = 0 \,.
\end{equation}
In the super-horizon regime, the gradient term can be neglected and we obtain $a^3 \dot{\zeta}_c m^2_{\sigma, {\rm eff}}/m_\sigma^2 = {\rm cst}$, equivalent to $\dot{\zeta}_c m^2_{\sigma, {\rm eff}} = 0$ as $a\to+\infty$. This yields the same two solutions as above, namely either: (i) $\sigma=0$ and hence the conservation of $\zeta_c$, or (ii) $m^2_{\sigma, {\rm eff}}=0$. 

\vskip 4pt 
This completes the proof of the consistency between the single-field EFT and super-horizon regimes.

\section{Non-symmetric mass parametrization}
\label{app2}
\noindent
A central assumption throughout this paper is the use of the symmetric mass parametrization~\eqref{eq: masses parametrization}, which effectively collapses both linear parameters $m_\sigma^2$ and $\eta_\perp$ onto a single one: $\lambda$. In this appendix, we provide additional details on the bispectrum shape for a non-symmetric mass parametrization. To this end, we introduce a new parameter $\Delta\lambda \in [-1, +1]$, such that
\begin{equation}
    m_\sigma^2 = -\lambda^2(1+\Delta\lambda)H^2 \,, \quad m_{\sigma, \text{eff}}^2 = +\lambda^2(1-\Delta\lambda)H^2 \,.
\end{equation}
The relative difference $m_{\sigma, \text{eff}}^2-m_\sigma^2=2\lambda^2H^2$ remains independent of $\Delta\lambda$, and the bare mass $m_\sigma^2$ is still negative while the effective one $m_{\sigma, \text{eff}}^2$ is positive. The case $\Delta\lambda>0$ (resp.~$\Delta\lambda<0$) increases (resp.~decreases) the strength of the tachyonic instability, while for $\Delta\lambda=0$ we recover the symmetric mass parametrization~\eqref{eq: masses parametrization}. As an example, angular inflation studied in Sec.~\ref{sec: Concrete realization} corresponds to $\lambda^2=2\eta_\perp^2$ and $\Delta\lambda=+0.5$. For a heavy entropic field, the dynamics of linear fluctuations for the non-symmetric mass parametrization can be described by an effective single-field theory with the imaginary sound speed:
\begin{equation}
    c_s^{-2} = \frac{\Delta\lambda-1}{\Delta\lambda+1}<0 \,.
\end{equation}
Notice that $\Delta\lambda\to+1$ corresponds to $|c_s|\to\infty$ so that the tachyonic instability is infinitely strong. This limit should, however, be understood as purely formal: as $\Delta\lambda$ approaches $+1$, the single-field effective description ceases to be reliable and cannot be trusted.
\begin{figure}[t!]
    \centering
    \includegraphics[width=8.5cm]{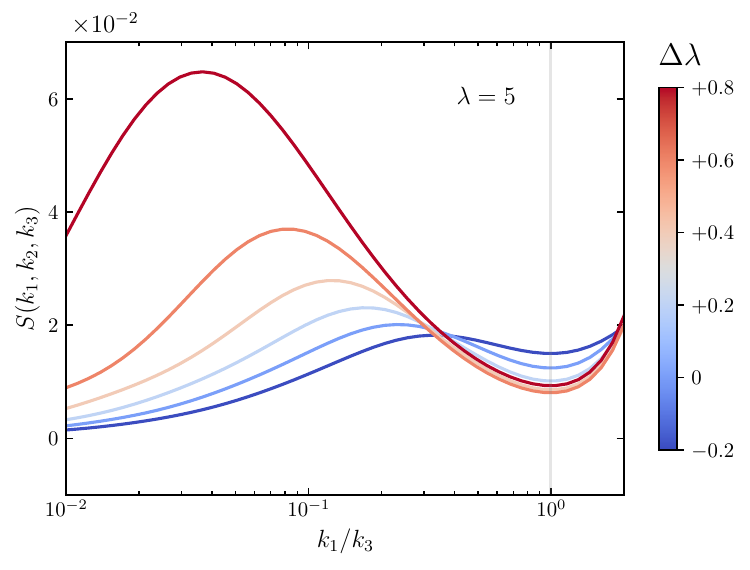}
    \caption{\it Dimensionless bispectrum shape $S(k_1, k_2, k_3)$ in the isosceles configuration $k_2=k_3$ from the folded limit $k_1/k_3=2$ to the squeezed limit $k_1/k_3\to0$ for the cubic interaction $\sigma^3$, fixing $\lambda=5$, and varying $\Delta\lambda$. We have not normalized the shape to unity in the equilateral configuration (however we have set the cubic Wilson coefficient to unity). The case $\Delta\lambda=+1$ sets the entropic field effective mass to zero (this is ultralight regime~\cite{Achucarro:2016fby}), while the case $\Delta\lambda\to-1$ approaches the strongly-mixed regime where the entropic effective mass is almost entirely dominated by the turning rate.}
    \label{fig: non-symm mass param}
\end{figure}
\vskip 4pt
For concreteness, we numerically compute the bispectrum generated by the cubic interaction $\sigma^3$ and display its shape for isosceles kinematic configurations in Fig.~\ref{fig: non-symm mass param}.\footnote{We have checked that the results are qualitatively similar for other cubic interactions.} For positive $\Delta\lambda\gtrsim0$, we continue to observe both an enhanced folded limit and a resonance in mildly squeezed configurations. As the instability becomes stronger, the resonance grows in amplitude and shifts toward increasingly squeezed configurations. These results further support the universality of the bispectrum produced by a transient tachyonic instability: {\it its robust observational signature is the simultaneous presence of a tachyonic resonance and a magnified folded limit}, while quantitative features such as the resonance position and amplitude depend on the underlying microphysics. The qualitative behavior is therefore largely insensitive to the symmetric mass parametrization adopted throughout this work. Conversely, as $\Delta\lambda\to-1$, the tachyonic instability gradually disappears, together with these characteristic signatures. Interestingly, the enhancement of the folded limit is essentially independent of the mass parametrization.

\bibliographystyle{apsrev4-1.bst}
\bibliography{Refs.bib}
\end{document}